\documentclass[twocolumn,showpacs,preprintnumbers,amsmath,amssymb]{revtex4}
\usepackage{graphicx}
\newcommand{\ket}[1]{|#1\rangle}
\newcommand{\bra}[1]{\langle#1|}
\newcommand{\abs}[1]{|#1|}
\newcommand{\twointeg}[4]{\{#1,#2|#3,#4\}}

\begin{document}
\title{Structure-dependent exchange in the organic magnets Cu(II)Pc and Mn(II)Pc}
\author{Wei Wu}\email{wei.wu@ucl.ac.uk}
\author{A.\ Kerridge}
\altaffiliation[Present address: ]{UCL Department of Chemistry, 20 Gordon Street, London, WC1H 0AJ, U.K.}
\author{A. H.\ Harker}
\author{A. J.\ Fisher}\email{andrew.fisher@ucl.ac.uk}
\affiliation{UCL Department of Physics and Astronomy and London Centre for Nanotechnology,
University College London, Gower Street, London WC1E 6BT, U.K.}

\begin{abstract}
We study exchange couplings in the organic magnets copper(II) phthalocyanine (Cu(II)Pc) and manganese(II) phthalocyanine (Mn(II)Pc) by a combination of Green's function perturbation theory and \textsl{ab initio} density-functional theory (DFT). Based on the indirect exchange model our perturbation-theory calculation of Cu(II)Pc qualitatively agrees with the experimental observations. DFT calculations performed on Cu(II)Pc dimer show a very good quantitative agreement with exchange couplings that our theoretical group extracts by using a global fitting for the magnetization measurements to a spin-$\frac{1}{2}$ Bonner-Fisher model. These two methods give us remarkably consistent trends for the exchange couplings in Cu(II)Pc when changing the stacking angles. The situation is more complex for Mn(II)Pc owing to the competition between super-exchange and indirect exchange.
\end{abstract}

\pacs{71.10.-w,71.15.Mb,71.35.Gg,71.70.Gm,75.10.Pq,75.50.Xx}

\maketitle

\section{Introduction}\label{sec:introduction}
Recently molecular spintronics \cite{Kinoshita91,Maniero00,Datta04,rocha05,pramanik2007} has become a very active
inter-disciplinary topic. This is because localized spins in molecular complexes can have very long spin relaxation times (up to of order one second) \cite{pramanik2007}, while the chemical engineering of such complexes is much more flexible than is the case in conventional inorganic-semiconductor electronics. The main building blocks of molecular spintronics, namely radicals containing localized electrons, are promising candidates both for spintronics and for quantum information processing. Against this background, experimental and theoretical studies of the
magnetism in spintronics-related organic materials are crucial
for the development of devices such as molecular magnetic random-access memory (MRAM) \cite{Engel05,emberly2002,pati2003}.

\begin{figure}
\begin{tabular}{c}
\includegraphics[width=6cm,height=2.5cm]{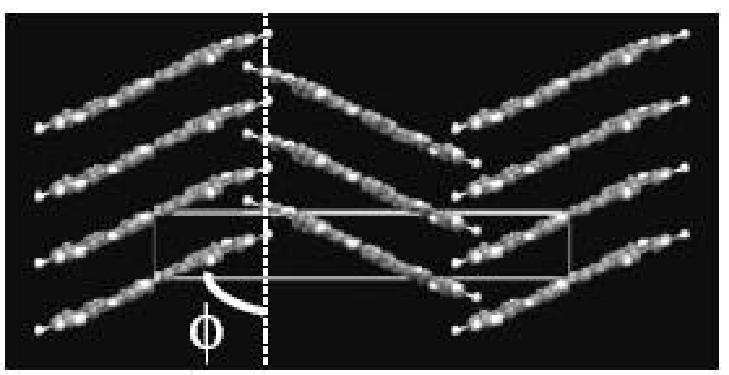}\\
(a)\\
\includegraphics[width=6cm,height=2.5cm]{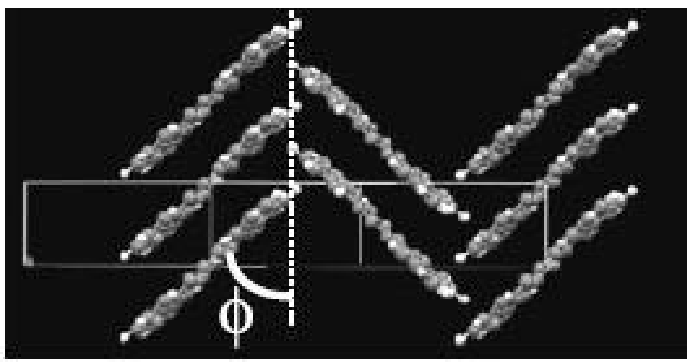}\\
(b)\\
\end{tabular}
\caption{Schematic arrangements of (a) the $\alpha$- and (b) $\beta$-phase MPc structures. The stacking angle is $\phi$ in each case.}\label{pic:stacking}
\end{figure}

\begin{figure}
\begin{tabular}{c}
\includegraphics[width=6cm,height=6cm]{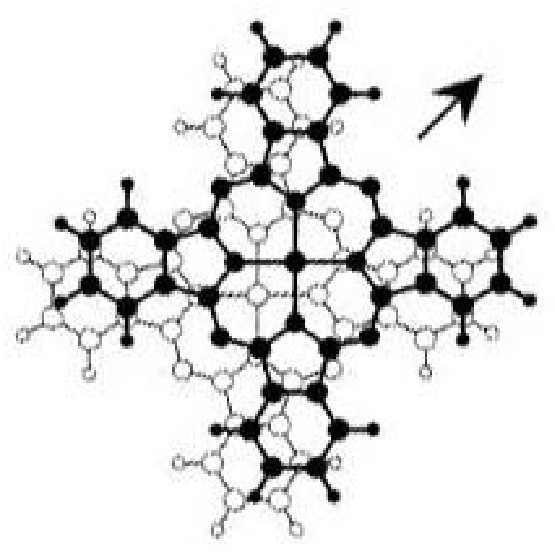}\\
(a)\\
\includegraphics[width=6cm,height=6cm]{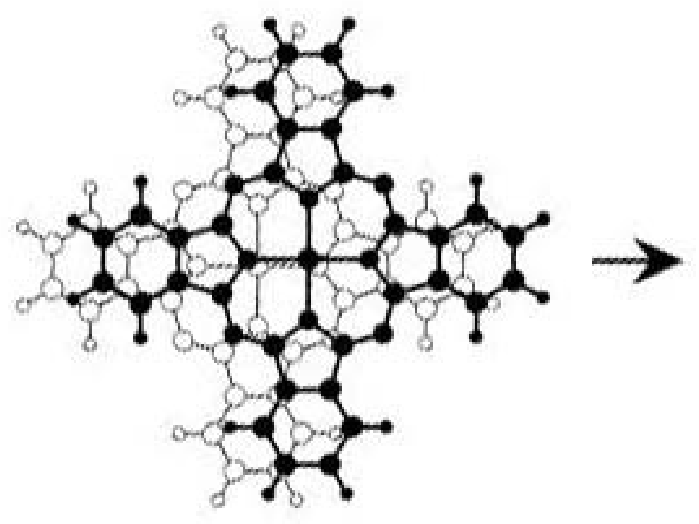}\\
(b)\\
\end{tabular}
\caption{Two models for stacking orientations for $\alpha$-Cu(II)Pc: (a) "$\times$"; (b)"$+$".}\label{pic:modelab}
\end{figure}

There is a long history of research on metal phthalocyanines (MPc), because of their commercial applications and excellent electro-optical properties \cite{porphyrinbooks}. In particular, their magnetic properties have been extensively studied \cite{cgb70,yamada98,evangelisti02}. Mn(II)Pc was one of the first molecular magnets \cite{mitra83}; its properties were shown to depend critically on the stacking of the planar molecular $\pi$-systems.  We define a stacking angle as shown in Figure~\ref{pic:stacking}; the $\beta$-Mn(II)Pc crystal (stacking angle $45^{\circ}$) was found to be ferromagnetic, while the $\alpha$-Mn(II)Pc thin film (stacking angle
$65^{\circ}$) was shown to be antiferromagnetic. However, there is still debate about the details of the molecular stacking in the $\alpha$-phase Cu(II)Pc material: transmission-electron diffraction (TED) observations for Cu(II)Pc on a KCl (001) surface \cite{ashita} suggested that the stacking orientation in the $\alpha$-phase was the so-called "$\times$" model as shown in Figure~\ref{pic:modelab}(a). However, the most recent TED experiments  \cite{hoshino} by contrast indicated the orientation is the "$+$" model shown in Figure~\ref{pic:modelab}(b). The arrows in Figure~\ref{pic:modelab} show the direction of the displacement of the Pc molecules between successive layers; these directions differ by a rotation of $45^\circ$. Since this issue is not yet resolved, and given that the substrate used in \cite{ashita, hoshino} differs from that used in \cite{sandrine}, in the following discussion we adopt the "$+$" model.

Recently Heutz \cite{sandrine}, \textsl{et
al.}, have performed further magnetic measurements on different
phases of Cu(II)Pc and Mn(II)Pc by using SQUID magnetometry. In the remainder of this paper, we describe these spin systems by using a Heisenberg spin-chain model \cite{heisenberg}, which is believed to be a good description for organic systems containing localized spin centers:
\begin{equation}
\hat{H}_{\mathrm{eff}}=-2J\sum{\mathbf{S}_i\cdot\mathbf{S}_j}.
\end{equation}
Note that with the sign convention we adopt, a positive exchange constant $J$ corresponds to ferromagnetic coupling, while a negative $J$ describes anti-ferromagnetic interactions.
In these experiments Mn(II)Pc powder samples ($\beta$-phase) show strongly ferromagnetic (FM) coupling with
$J\approx 11.45 \mathrm{K}$, while Mn(II)Pc films ($\alpha$-phase) grown on an inert Kapton substrate shows a relatively much weaker anti-ferromagnetic (AFM) coupling with $J\approx -1.61\mathrm{K}$.
Cu(II)Pc powder ($\beta$-phase) is found to be very weakly ferromagnetic (indeed nearly paramagnetic) with $J\approx 0\mathrm{K}$, but Cu(II)Pc films ($\alpha$-phase) and Cu(II)Pc whose growth is templated by a layer of 3,4,9,10-perylenetetracarboxylic dianhydride (PTCDA) pre-deposited on the Kapton substrate are found to be more strongly anti-ferromagnetic ($J\approx -1.50 \mathrm{K}$). The exchange constants for Mn(II)Pc are extracted from the intercept of the inverse susceptibility versus temperature, while those for Cu(II)Pc are found by a global fit of the experimental data to a finite $S=1/2$ Heisenberg spin chain model (the so-called Bonner-Fisher model \cite{bonner}). This model is expected to be sufficiently accurate, despite its neglect of inter-chain couplings, provided the temperature is not too low relative to the exchange constants.

These experiments clearly show the switching of magnitudes and signs of the exchange couplings as the molecular packing varies from phase to phase, and also that the magnetic properties are determined by the structure ($\alpha$ versus $\beta$), not by the sample morphology (powder
versus thin film). The results confirm the previously measured
\cite{cgb70,yamada98} difference between the $\alpha$ and $\beta$
phases of Mn(II)Pc, and also show that a corresponding difference
exists for Cu(II)Pc, although in this case the $\beta$ phase is
paramagnetic rather than ferromagnetic.

Despite the long history of experimental work on MPc, there have been very few systematic
theoretical studies of the mechanisms underlying the variation in the exchange interactions; the problem is complicated by the molecular structure and rather
weak spin-spin interactions compared to conventional inorganic
semiconductors. In this paper we aim to gain both a picture of the physics driving the
structure-dependent exchange, and a quantitative
understanding of its magnitude, in Cu(II)Pc and Mn(II)Pc.

Our remaining discussion falls into five sections. In \S\ref{sec:overview}, we introduce the different mechanisms for exchange and describe state-of-art quantitative methods to evaluate exchange interactions by using DFT and the broken symmetry concept. We also describe the atomic and electronic structure of the systems we consider.  In \S\ref{sec:greenfunction} we perform Green's-function perturbation theory calculations for exchange interactions to get a rough picture of essential physics. In \S\ref{sec:dft} we use DFT with the hybrid exchange-correlation functional B3LYP to evaluate the exchange interactions quantitatively. At the end we draw our conclusions in \S\ref{sec:conclusion}.

\section{Theoretical Overview}\label{sec:overview}
\subsection{Qualitative description of exchange interactions between localized electrons}\label{sec:exchange}

The various processes contributing to the exchange interactions between localized spins were extensively considered by Anderson \cite{anderson}. His original paper considered exchange interactions between magnetic ions in ionic crystals, but the same concepts and arguments apply in the case of the bi-radical studied here. The \textit{direct} exchange interaction originates from a quantum exchange term of the Coulomb interaction between localized electrons, e.g., $d$-electrons in a magnetic ion; this always gives rise to ferromagnetic exchange interactions. However, for MPc the direct exchange interaction can generally be neglected owing to the large molecular size and the localization of the metal electrons.

The \textit{super}-exchange arises from the terms in the Hamiltonian that tend to delocalize electrons. It is then necessary to take the hopping perturbation to higher order in order to reach an excited state in which an electron is transferred onto a neighboring magnetic site.  For example, in a simple Hubbard model the second order in perturbation theory leads to an exchange interaction $-2t^2/U$, where $t$ is the transfer integral between sites and $U$ is the on-site Coulomb repulsion. In a more realistic model there are many more super-exchange pathways but the same principles remain: according to the definition we adopt, for a process to be classified as super-exchange it should proceed through virtual states involving the migration of charge between the magnetic centers.  The super-exchange mechanism is
especially important when the magnetic atoms are separated by
non-magnetic species, for example inorganic anions or organic
ligands, through which charge can pass from one magnetic atom to another.
We should note that the super-exchange vanishes not only when the distances are large but also when the  the transfer
of electrons between the magnetic centers and these covalent "bridges" is symmetry-forbidden.

The indirect exchange interaction between electronic spins is
similar to the indirect exchange interaction between nuclear
moments, which is mediated by the conduction (or valence) electrons: a polarization of the conduction electrons around one local moment is propagated to another, giving rise to an effective interaction.  This nuclear interaction was first discovered by Ruderman and Kittel \cite{rk}, and independently in molecular physics by Ramsey and Purcell \cite{ramsey53,ramsey53b} and by Bloembergen and Rowland \cite{br}; the generalization to localized electronic moments is due to Kasuya \cite{kasuya} and Yosida \cite{yosida}. In metals it leads to a long-range exchange coupling that is oscillatory in sign.

We can sharpen the distinction between the different types of exchange by writing the full electronic Hamiltonian as
\begin{equation}
\hat{H}=\hat{H}_\mathrm{spin}+\hat{H}_\mathrm{ligand}+\hat{V}_\mathrm{direct}+\hat{V}_\mathrm{hop}+\hat{V}_\mathrm{polarize}.
\end{equation}
The terms are defined as follows.  $\hat{H}_\mathrm{spin}$ corresponds to the isolated spins, and $\hat{H}_\mathrm{ligand}$ to the rest of the material (decoupled from the spins), both being taken to include an appropriate (spin-independent) mean field. We take $\hat{H}_0=\hat{H}_\mathrm{spin}+\hat{H}_\mathrm{ligand}$ to be our unperturbed Hamiltonian.  $\hat{V}_\mathrm{direct}$ then involves the direct Coulomb interaction, coupling the spins to one another without involving the ligands; $\hat{V}_\mathrm{hop}$ involves all processes that transfer an electron between the magnetic species and the ligands; while $\hat{V}_\mathrm{polarize}$ includes all other (non-charge-transferring) interactions between the magnetic species and the ligands.

We can then develop a perturbation expansion for the full Green's function $G$ (as, for example, in \cite{ourpaper2}) as \begin{equation}
G=G_0+G_0VG_0+G_0VG_0VG_0+\ldots
\end{equation}
where $V$ is the perturbation $\hat{V}_\mathrm{direct}+\hat{V}_\mathrm{hop}+\hat{V}_\mathrm{polarize}$ and $G_0$ is the Green's function corresponding to unperturbed Hamiltonian.  The effective exchange is then recovered by computing an effective Hamiltonian
\begin{equation}
E-\hat{H}_\mathrm{eff}=G^{-1}
\end{equation}
within a ground-state manifold of configurations differing only in the spin orientations.
Within this picture, the direct exchange corresponds to the first-order term involving $\hat{V}_\mathrm{direct}$, while super-exchange and indirect exchange correspond to higher-order terms involving $\hat{V}_\mathrm{hop}$ and $\hat{V}_\mathrm{polarize}$ respectively.  It is clear from the definitions of the various terms in $\hat{V}$ that the virtual states that couple to the ground-state manifold are orthogonal; therefore (at least to low orders in $\hat{V}$) we do not need to consider cross-terms between the different operators.

In this paper we will neglect $\hat{V}_\mathrm{direct}$, for the reasons given above.  Appropriate expressions for $\hat{V}_\mathrm{hop}$ and $\hat{V}_\mathrm{polarize}$ are given in \S\ref{sec:greenfunction} below.

\subsection{Quantitative calculation: DFT calculations of exchange
couplings in bi-radicals}
Density Functional Theory (DFT) \cite{bagus,ziegler,noodleman,yamaguchi,matteo,martin,illas00} is a powerful tool for accurate prediction of the exchange interactions in chemically complex wide-gap materials.  However current density functionals, based on Kohn-Sham theory, give a poor representation of singlet states containing a pair of localized electron spins since the Kohn-Sham orbitals are constrained to respect the symmetry of the system and are therefore generally formed from linear combinations of the single-center wavefunctions.  One has to use instead a
so-called `broken-symmetry' method \cite{noodleman}, in which the
magnetic orbitals are localized in different radical centers, with
their spins oppositely-aligned. Recently, Martin and Illas \cite{martin,illas00} checked the performance of different exchange-correlation functionals in the calculation of
magnetic couplings, and found the choice of exchange functionals
is extremely important, while the role of the correlation
functional is minor. Although the precise reasons are unclear, it is empirically found that it is necessary to mix some proportion of exact exchange into the functional in order to obtain results that agree with experiment or with higher-quality quantum chemistry results for small molecules---crudely, this may be understood as requiring some balance between the over-localization of electrons in a Hartree-Fock calculation, and the excessive delocalization in standard density functionals.  In particular the B3LYP functional \cite{b3lyp,becke,lyp}, which mixes about one quarter Hartree-Fock exchange, has been found to give good results for di-nuclear molecules, organic bi-radicals, and spins localized at defects in carbon-containing materials \cite{martin,illas00,chan04}. In \S\ref{sec:dft}, we perform DFT with B3LYP exchange-correlation functional to
calculate exchange interactions in Cu(II)Pc dimers based on this broken-symmetry concept.

Although the DFT and perturbative approaches to the problem appear at first sight to be quite different, one can think of them as representing in different ways the same response of the electronic system to its spin-dependent interactions.  In the case of the DFT, this response is represented as a change in the Kohn-Sham states, while in the model approaches the many-electron wave function responds by including small admixtures of excited-state configurations.

\subsection{The electronic structure of isolated Cu(II)Pc and Mn(II)Pc}
\label{sec:elstruc}

Before we perform the perturbation theory calculation for the exchange interaction, we need to understand the nature of the one-electron states in the isolated molecules. We used the Gaussian 98 code \cite{gaussian98}, performing a DFT calculation with the B3LYP \cite{b3lyp} exchange-correlation functional and a 6-31G \cite{ditchfield} basis set, to optimize the molecular geometry of isolated Cu(II)Pc and Mn(II)Pc molecules. We then use the key Kohn-Sham states emerging from DFT calculation which are nearest to the HOMO-LUMO gap as a basis to perform a separate perturbation-theory calculation.

\begin{figure}[htbp]
\begin{tabular}{c}
\includegraphics[width=7cm,height=7cm]{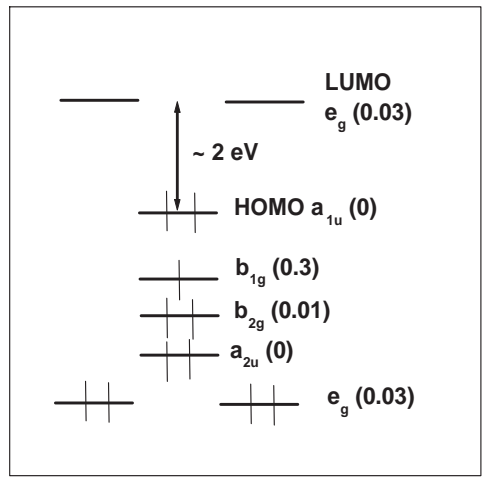}\\
(a)\\
\includegraphics[width=7cm,height=7cm]{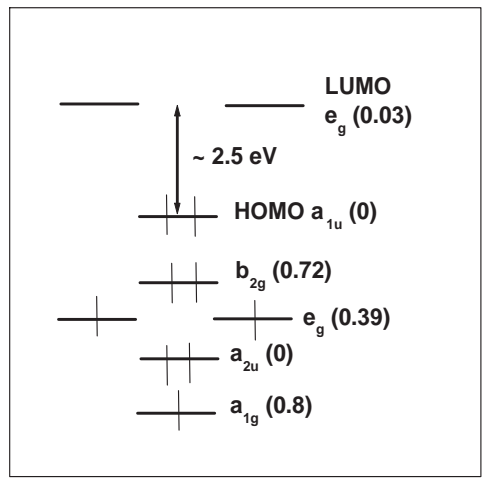}\\
(b)\\
\end{tabular}
\caption{The schematic of the key states from our Gaussian DFT calculations of the electronic structures of
 isolated (a) Cu(II)Pc and (b) Mn(II)Pc. The majority-spin contributions to the Mulliken charge on the transition-metal
atoms are shown in brackets (the minority-spin contributions for the doubly-occupied states are similar).}\label{pic:gausscupcmnpc}
\end{figure}

Our level scheme for Cu(II)Pc is shown in Figure~\ref{pic:gausscupcmnpc}(a). The states are labeled by the irreducible representations of the $D_{4h}$ point group.  From the Mulliken population analysis we can identify $b_{1g}$ as a metal d-orbital which is hybridized with the Pc ring (Mulliken
charge 0.30). The total Mulliken
charge on copper is +0.97. The total Mulliken spin density on the copper atom is $0.68$; this is consistent
with the existence of one singly-occupied orbital, with a spin
mainly but not entirely localized on the copper atom. The overall symmetry of the electronic state is $^2B_{1g}$. We found that the occupied molecular orbital with the largest
Kohn-Sham eigenvalue is not the singly-occupied $b_{1g}$ state,
but the $a_{1u}$ state; this is slightly different from the early extended
H\"{u}ckel calculations \cite{zernergouterman}. This highlights the importance
of two-electron Coulomb terms in determining the configuration:
doubly occupying the $b_{1g}$ state would incur a large Coulomb
penalty because the charges would spend much of their time
localized in the Cu 3d states, whereas the double-occupancy
penalty for the more diffuse $a_{1u}$ state is much smaller.

For Mn(II)Pc, the expected total spin is $S=3/2$ \cite{cgb70, sandrine}.  Our Gaussian calculation gave the overall electronic
configuration $^4A_{1g}$, with three singly-occupied one-electron
levels having $a_{1g}$ and $e_g$ (twice) symmetry. The total
Mulliken charge on Mn is +1.14; again this is of the same order as,
but somewhat less than, the nominal +2 valence. The Mulliken spin
density on the Mn atom is $+3.1$. Our calculation results are similar to those in \cite{zernergouterman} which give $^4A_{2g}$ total symmetry and the same single-occupied states. However, this picture of the electronic structure is not the only one. Liao \textit{et.\ al}  \cite{liao} also used DFT methods and found an
electronic configuration in which the three unpaired electrons
occupy the $a_{1g}$, $b_{2g}$, and one $e_{g}$ state while the
other $e_g$ state is doubly occupied, to give an electronic symmetry
$^4 E_g$. This calculation is in agreement with the more recent
magnetic circular dichroism (MCD) and UV-vis measurements
\cite{williamson92} of the molecule in an argon matrix, but
differs from the early magnetic measurements of solid Mn(II)Pc.  It
is possible that the $^4 E_g$ configuration (which would lead to a Jahn-Teller distortion, because of its orbital degeneracy) may be favored in the
isolated molecule or in the argon matrix, with the $^4A_{2g}$
state favored in the bulk material (where no Jahn-Teller distortion has been observed).

\section{Green's function perturbation calculations}\label{sec:greenfunction}

\subsection{Super-exchange calculation}
We aim to understand the mechanism of exchange couplings between
neighboring Mn(II)Pc and Cu(II)Pc molecules observed in experiments
\cite{cgb70, yamada98, sandrine}.  We first consider the super-exchange
contributions.

\subsubsection{Cu(II)Pc}
As explained in \S\ref{sec:overview}, the super-exchange
contribution is generally dominant when considering the exchange interaction
between localized spins in insulating materials. However, Cu(II)Pc is
an example of a situation where this interaction is expected to be
negligible.  This is because the unpaired spin is located in a
$b_{1g}$ orbital \cite{zernergouterman}, but there is no low-energy state of $b_{1g}$
symmetry in the ligand available to hybridize with it.  Therefore,
as long as the symmetry of the molecule remains $D_{4h}$ (believed
to be an excellent approximation even in the crystal) the
spin-carrying electron is ``tied'' to the Cu site and no
super-exchange can take place except by direct hopping from the Cu
orbitals onto the neighboring molecule.  The amplitude for this process is expected to be very small.

\subsubsection{Mn(II)Pc}\label{sec:superexmn}
\begin{figure}[htbp]
\begin{center}
\includegraphics[width=9cm,height=6.5cm]{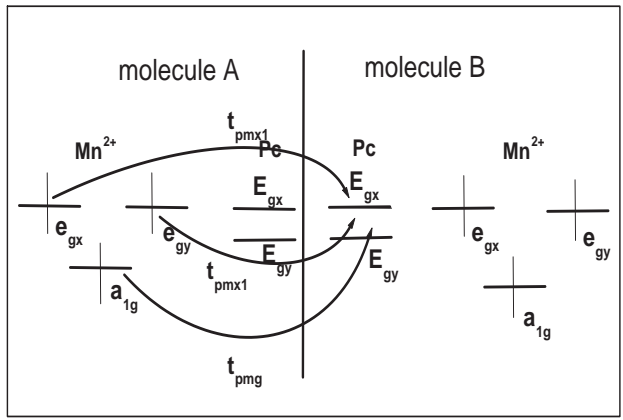}
\end{center}
\caption{The possible inter-molecular
transitions.}\label{pic:temnpc}
\end{figure}

In the case of Mn(II)Pc, there are three unpaired electrons per molecule occupying the  $a_{1g}$, $e_{gx}$, and $e_{gy}$ molecular orbitals (see \S\ref{sec:elstruc}). We can call these metal states because these orbitals originate from the splitting of atomic $3d$-states in the molecular environment which has $D_{4h}$ symmetry.  The two $S=3/2$ spins of the individual molecules can be combined to form
a total spin of
$0,1,2,3$. We therefore need, in principle, three independent
parameters in the spin Hamiltonian to characterize fully the
relative energies of these states. If we neglect spin-orbit
coupling, the Hamiltonian must be invariant under simultaneous
rotations of both spins and therefore must have the form
\begin{eqnarray}\label{eq:threehalveswiththreehalves}
\hat{H}=-J_1(\mathbf{S}_{A}\cdot\mathbf{S}_{B})-J_2(\mathbf{S}_{A}\cdot\mathbf{S}_{B})^2-
J_3(\mathbf{S}_{A}\cdot\mathbf{S}_{B})^3,
\end{eqnarray}
where $J_1, J_2$, and $J_3$ are exchange couplings and $A$, $B$ label the molecules. We can find all
three parameters from the $4\times 4$ Hamiltonian matrix spanned by the states
with total $z$ spin angular momentum $M_S=0$: $\ket{(3/2)_{A},(-3/2)_{B}},
\ket{(1/2)_{A},(-1/2)_{B}}, \ket{(-1/2)_{A},(1/2)_{B}}$, and
$\ket{(-3/2)_{A},(3/2)_{B}}$. We use a similar method to that described in \cite{ourpaper2}: we construct the effective Hamiltonian based on an extended Hubbard model by Green's-function perturbation theory, compare this with equation~(\ref{eq:threehalveswiththreehalves}),
and extract the exchange
constants. For simplicity we include only
intermediate states where a single electron is transferred between adjacent Mn ions via the ligand $e_{g}$
states (i.e. we neglect the possibility that two or more electrons transfer together).  We also neglect direct electron transfer between the d states of the Mn ions, because these states are quite well localized. Our extended Hubbard model reads:
\begin{eqnarray}
\hat{H}&=&\hat{H}_0+\hat{H}_t+\hat{H}_p;
\\\hat{H}_0&=&\{\sum_{i} E_i\hat{n}_{i}
\\\nonumber&&+v_{pmn}(\hat{n}_{E_{gx}A}+\hat{n}_{E_{gy}A})(\hat{n}_{e_{gx}A}+\hat{n}_{e_{gy}A}+\hat{n}_{a_{1g}A})
\\\nonumber&&+u_{gx}\hat{n}_{a_{1g}A}(\hat{n}_{e_{gx}A}+\hat{n}_{e_{gy}A})
\\\nonumber&&+u_{xx}(\hat{n}_{e_{gx}A\uparrow}\hat{n}_{e_{gx}A\downarrow}+\hat{n}_{e_{gy}A\uparrow}\hat{n}_{e_{gy}A\downarrow})
\\\nonumber&&+u_{gg}\hat{n}_{a_{1g}A\uparrow}\hat{n}_{a_{1g}A\downarrow}\}
\\\nonumber&&+\{A\Leftrightarrow B\};
\\\hat{H}_t&=& \{\sum_{\sigma}[t_{pmg}\hat{c}_{(a_{1g}A)\sigma}^{\dagger}(\hat{c}_{(E_{gx}B)\sigma}+\hat{c}_{(E_{gy}B)\sigma})
\\\nonumber&&+t_{pmx1}(\hat{c}_{(e_{gx}A)\sigma}^{\dagger}\hat{c}_{(E_{gx}B)\sigma}+\hat{c}_{(e_{gy}A)\sigma}^{\dagger}\hat{c}_{(E_{gy}B)\sigma})
\\\nonumber&&+t_{pmx2}(\hat{c}_{(e_{gx}A)\sigma}^{\dagger}\hat{c}_{(E_{gy}B)\sigma}+\hat{c}_{(e_{gy}A)\sigma}^{\dagger}\hat{c}_{(E_{gx}B)\sigma})+h.c.]\}
\\\nonumber&&+\{A\Leftrightarrow B\};
\\\hat{H}_p &=& -K\mathbf{s}\cdot\mathbf{S}.
\end{eqnarray}
Here $a_{1g},e_{gx}$, and $e_{gy}$ label the metal states; $E_{gx}$ and $E_{gy}$ label LUMO ligand states for distinguishing metal and ligand $e_g$ states \cite{porphyrinbooks}. $\hat{H}_0$ includes the single-particle energies $E_i$ where $i$ runs through $a_{1g},e_{gx},e_{gy},E_{gx},E_{gy}$, the Coulomb interaction between metal and ligand states, and the on-site Coulomb interactions. There are two parts in the perturbation: one is $\hat{H}_t$ which transfers electrons between molecules and the other is $\hat{H}_p$ representing the interaction between $\frac{1}{2}$-spin ($\mathbf{s}$) in the ligand and $\frac{3}{2}$-spin ($\mathbf{S}$) on the metal within the molecule.  We suppose that $\hat{H}_p$ is itself ultimately a representation of a further super-exchange processes, and therefore like $\hat{H}_t$ originates in $V_\mathrm{hop}$ as defined in \S\ref{sec:exchange}.
$t_{pmg}$, $t_{pmx1}$ and $t_{pmx2}$ are the inter-molecular hopping integrals shown in Fig.~\ref{pic:temnpc}, $E_{g}$ and $e_{g}$ are the energies of ligand
and metal $e_g$ states relative to the energy level of $a_{1g}$ state, $u_{gx}$ is Coulomb interaction between the $a_{1g}$ and $e_{g}$ levels, $u_{xx}$ is the Coulomb interaction between two degenerate Mn $e_{g}$ states, and $v_{pmn}$ is the Coulomb interaction between the Mn and Pc states within a molecule.

From this Hamiltonian we can see when one electron
is transferred from the metal state of molecule A to the ring
state of molecule B where it can interact with the Mn spin; it is
through the interaction $\hat{H}_p$ that the spin projections $m_A$ and $m_B$
associated with the two molecules can change, thereby coupling the
four spin states: $\ket{(3/2)_{A},(-3/2)_{B}}$,
$\ket{(1/2)_{A},(-1/2)_{B}}$, $\ket{(-1/2)_{A},(1/2)_{B}},$ and
$\ket{(-3/2)_{A},(3/2)_{B}}$.  Note that in $D_{4h}$ symmetry, the $e_g$
states of Mn can hybridize effectively with the $e_g$ states of
the ring; the existence of unpaired spins in the $e_g$ states is
what makes super-exchange processes much more important in the
case of Mn(II)Pc.

There are 25 spatial configurations and each has four possible
spin states, giving a total of 100 states. We construct the
$100\times 100$ Hamiltonian matrix for $\hat{H}_0$ and $V$, and then extract the effective
Hamiltonian within the $4\times 4$ low-energy subspace by using
Green's function perturbation theory \cite{ourpaper2} to calculate the energy
shifts. By comparing this low-energy subspace with the
$\frac{3}{2}$-spin coupling matrix, we find that it can be written
in the form (\ref{eq:threehalveswiththreehalves}), with parameters
\begin{widetext}
\begin{eqnarray}
J_{1}&=&\frac{4}{3}K[\frac{t_{pmg}^2}{(-E_{g}+2u_{gx}-3v_{pmn})^2}+
\frac{t_{pmx1}^2}{(e_{g}-E_{g}+u_{gx}+u_{xx}-3v_{pmn})^2}\nonumber
\\&&+\frac{t_{pmx2}^2}{(e_{g}-E_{g}+u_{gx}+u_{xx}-3v_{pmn})^2}]+O(t^3);
\\J_2&=&-\frac{4}{9}K^2[\frac{t_{pmg}^2}{(-E_{g}+2u_{gx}-3v_{pmn})^3}+
\frac{t_{pmx1}^2}{(e_{g}-E_{g}+u_{gx}+u_{xx}-3v_{pmn})^3}\nonumber
\\&&+\frac{t_{pmx2}^2}{(e_{g}-E_{g}+u_{gx}+u_{xx}-3v_{pmn})^3}]+O(t^3);
\\J_3&=&\frac{4}{27}K^3[\frac{t_{pmg}^2}{(-E_{g}+2u_{gx}-3v_{pmn})^4}+
\frac{t_{pmx1}^2}{(e_{g}-E_{g}+u_{gx}+u_{xx}-3v_{pmn})^4}\nonumber
\\&&+\frac{t_{pmx2}^2}{(e_{g}-E_{g}+u_{gx}+u_{xx}-3v_{pmn})^4}]+O(t^3),
\end{eqnarray}
\end{widetext}

\normalsize
We note several features of this result.  First, the dominant
terms are those proportional to $t^2$, i.e. where electrons are
exchanged once between the molecules, as expected in a
super-exchange process. Second, the leading term in $J_1$ is
proportional to $Kt^2$, in $J_2$ to $K^2t^2$, and in $J_3$ to
$K^3t^2$; this is because $\hat{H}_{p}$ only couples states in
which $m_A$ and $m_B$ alter by one unit of angular momentum.
Finally, assuming the Coulomb energies are all large and positive,
$J_1$ is \textit{always} the same sign as $K$, irrespective of the
values of the various hopping terms. In general we expect that $K$, since it is dominated by super-exchange,
will be negative (corresponding to anti-ferromagnetic coupling in
our sign convention) and therefore $J_2$ will also lead to
anti-ferromagnetic coupling \textit{independent of the orientation of the molecules}.

Our conclusion about the failure of the super-exchange interaction
to change sign contrasts sharply with the explanation given by
Barraclough \textit{et al.}  \cite{cgb70} and by Yamada \textit{et
al.}  \cite{yamada98} for their experimental results, which they
ascribe to the competition between different super-exchange
pathways operating via nitrogen atoms.  However, this argument fails to take into account correctly the spin algebra---in particular, it ignores the fact that the three electron spins on each Mn atom are in fact tied together via strong intra-atomic Coulomb interactions, and so cannot be flipped independently.

\subsection{Indirect exchange calculation}

\subsubsection{Cu(II)Pc}\label{sec:indirectexcupc}

\begin{figure}[htbp]
\begin{center}
\includegraphics[width=9cm,height=6.5cm]{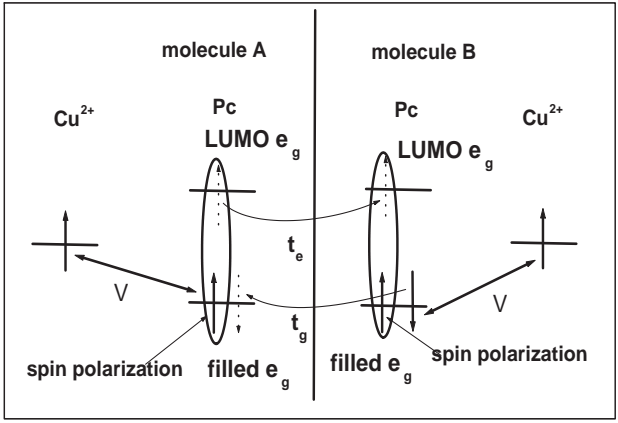}
\caption{Cu(II)Pc electron configuration and indirect exchange
scheme diagram. This scheme involves two filled $e_g$ states and
two empty $e_g$ states(LUMO).}\label{pic:cupcschme2}
\end{center}
\end{figure}

For the indirect exchange scheme in the Cu(II)Pc dimer (Figure ~\ref{pic:cupcschme2}), the unpaired electron spin on the metal polarizes the ligand by the two-body Coulomb
interaction; this spin polarization can transfer to the
neighboring molecule by orbital hybridization, and there interacts with the unpaired
spin of the neighboring molecule's metal ion.

Because the LUMOs are $e_g$ ligand states, we should
consider the filled ligand states with the same symmetry.
The two-body Coulomb interaction can be represented in second-quantized form as
\begin{eqnarray}\label{eq:coulombv}
\hat{v}=&&\sum_{\sigma\sigma^{\prime}}\sum_{ABDE}[\int{d\mathbf{r}d\mathbf{r}^{\prime}}\psi_{A}^{*}(\mathbf{r})
\psi_{B}^{*}(\mathbf{r}^{\prime})\frac{1}{\abs{\mathbf{r}-\mathbf{r}^{\prime}}}\psi_{D}(\mathbf{r}^{\prime})
\psi_{E}(\mathbf{r})]
\nonumber\\&&
\times
\hat{c}_{A,\sigma}^{\dagger}\hat{c}_{B,\sigma^{\prime}}^{\dagger}\hat{c}_{D,\sigma^{\prime}}\hat{c}_{E,\sigma},
\end{eqnarray}
where $\hat{c}$ and $\hat{c}^{\dagger}$ are the electron annihilation and
creation operators, and $A,B,D,E$ may each represent a metal orbital or a ligand
orbital.  Since we wish to consider processes in which the net charge of the metal ion does not change (i.e., contributions to $\hat{V}_\mathrm{polarize}$ rather than $\hat{V}_\mathrm{hop}$ in the language of \S\ref{sec:exchange}),
one of $(A,B)$ should correspond to a Cu state, i.e., metal $b_{1g}$
state and one to a Pc state, e.g., ligand $e_g$ state and
similarly for $(D,E)$.

Hence, overall the four indices may involve one entry for an $e_g$ LUMO state,
two entries for the $b_{1g}$ state, and one entry for a
doubly-filled ligand state: the highest-lying such states are
$a_{1u}$, $a_{2u}$, or $b_{2g}$ ligand states for single-molecule electronic structure \cite{porphyrinbooks}. However, because $a_{1u}$ and $a_{2u}$ are odd under inversion, but $b_{1g}$ and $e_{g}$ are
even, the two-electron integrals involving $a_{1u}$ and $a_{2u}$
are zero. Furthermore $B_{2g}$ transforms like $xy$ in $D_{4h}$
symmetry, $b_{1g}$ like $x^2-y^2$, and $e_{gx,y}$ like $zx,zy$.
The two-electron integral involving $b_{2g}$ is therefore odd in
either $y$ or in $x$ depending which $e_{g}$ state appears. So, in
fact the only important doubly-occupied states are the filled
$e_{g}$ states which appear slightly below the $a_{1u}$ and
$a_{2u}$.

\begin{figure}[htbp]
\begin{center}
\begin{tabular}{cc}
\includegraphics[height=4cm,width=4.5cm]{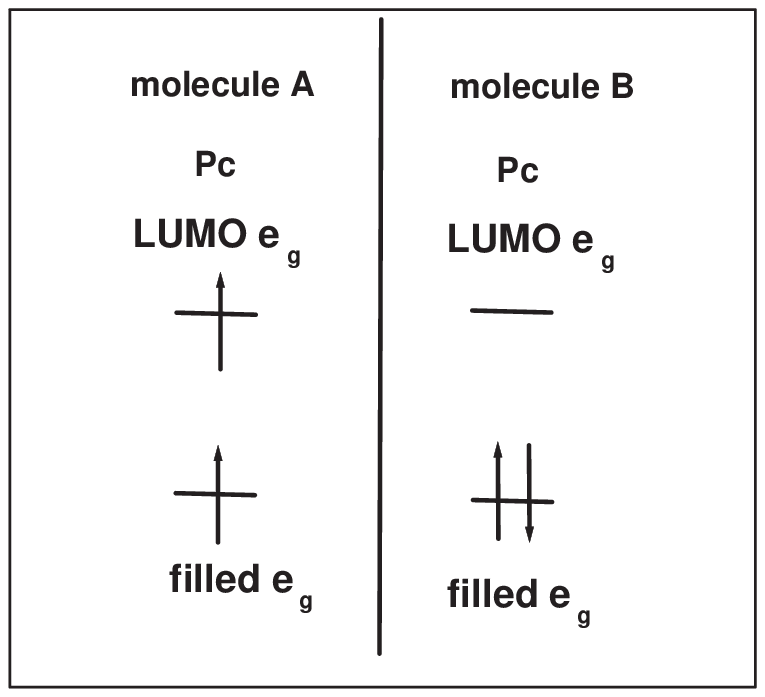}&
\includegraphics[height=4cm,width=4.5cm]{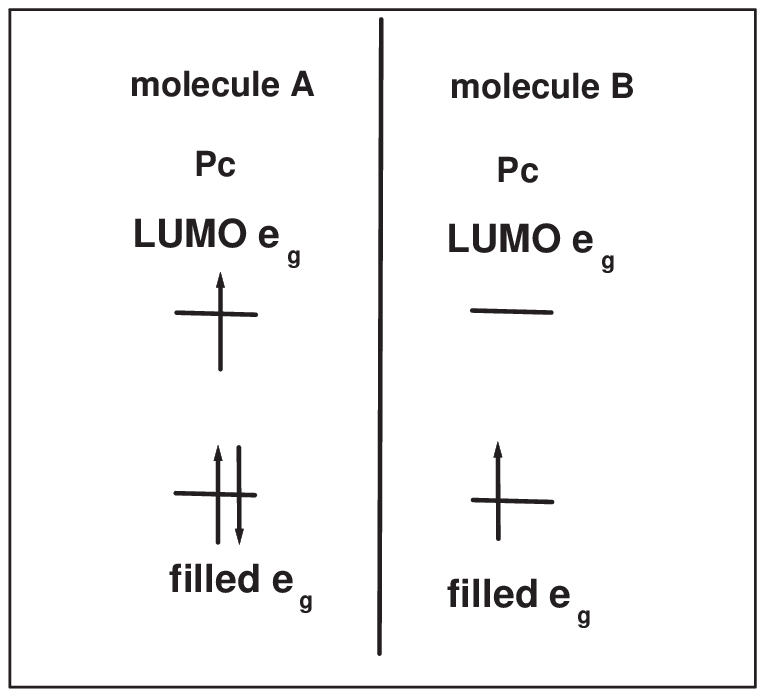}\\
(1)& (2)\\
\includegraphics[height=4cm,width=4.5cm]{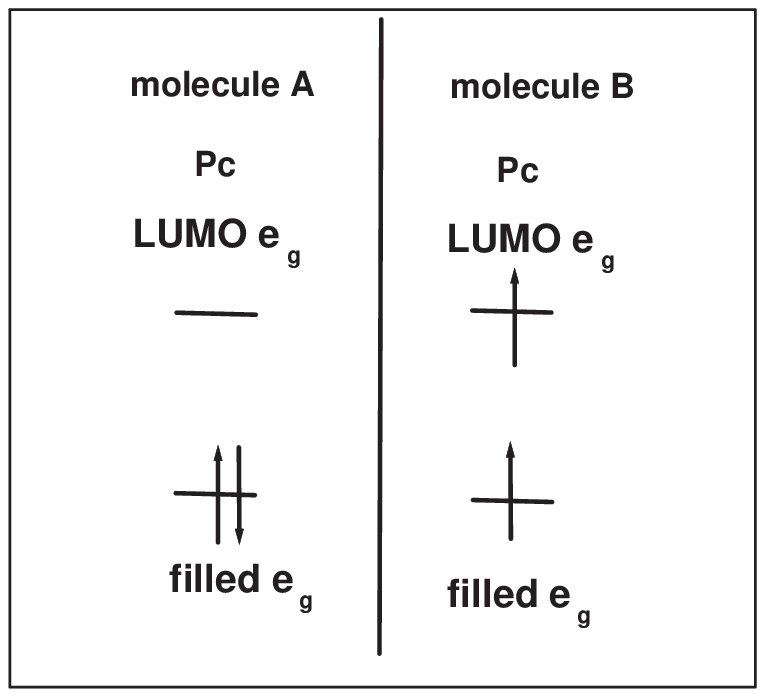}&
\includegraphics[height=4cm,width=4.5cm]{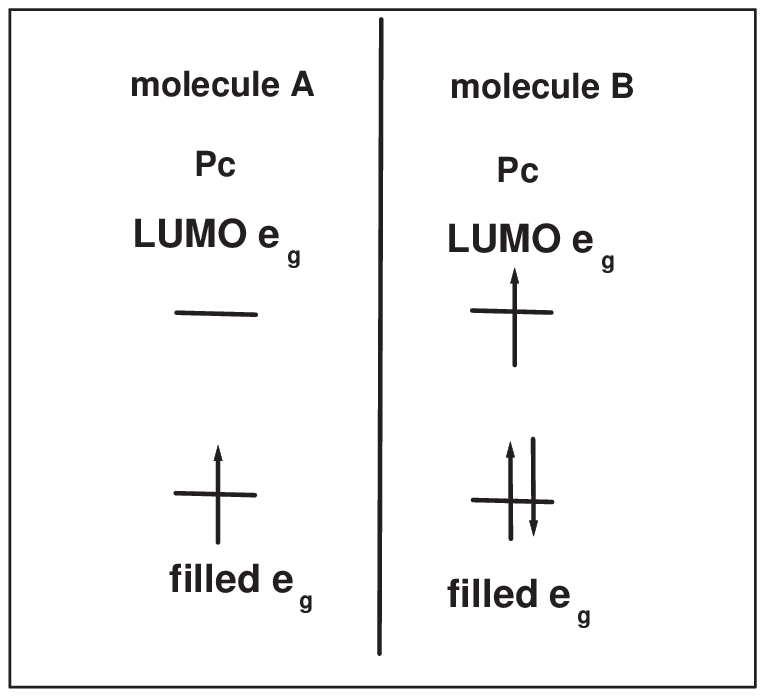}\\
(3)& (4)
\end{tabular}
\end{center}
\caption{The four states for calculating the transfer integrals between
polarized triplet states of different molecules as the example.}\label{pic:exciton4}
\end{figure}

In order to simplify the calculation we assume there is only one electron-hole pair produced in the Cu(II)Pc dimer (additional electron-hole pairs will cost more energy).
As in \S\ref{sec:superexmn}, we need consider only the situation where $M_{\mathrm{dimer}}^{\mathrm{total}}=M_{total}^{A}+M_{total}^{B}=0$ in order to extract the exchange constant. We find the Hamiltonian $\hat{v}$ can be written as the linear combination of the product of the metal spin operators and ligand spin-polarization operators owing to the preservation of total $S_z$ in the isolated molecule. We label the spatial LUMO state of the ligand by using "X", the filled states "G", and metal $b_{1g}$ state "b". We use the following notation for the two-electron integrals:
\begin{equation}
\twointeg{a}{b}{c}{d}=\int{d\mathbf{r}d\mathbf{r}^{\prime}a(\mathbf{r})^*b(\mathbf{r}^{\prime})^*
\frac{1}{\abs{\mathbf{r}-\mathbf{r}^{\prime}}}c(\mathbf{r})}d(\mathbf{r}^{\prime}).
\end{equation}
We can now apply Green's-function perturbation theory \cite{ourpaper2} to this problem; the perturbation includes the Coulomb interaction $\hat{v}$ which can polarize the spin in a ligand, and hopping $t$ that
transfers this polarization from one molecule to another. We find
that the leading term $J_1$ in the spin-$\frac{1}{2}$ couplings is given by:
\begin{eqnarray}\label{eq:excupc}
J_1&=&\frac{4\alpha^2t}{(U_g-U_x-E_x-2j_\mathrm{eh})^2};
\\\alpha&=&2\twointeg{X}{b}{b}{G};
\\U_g&=&\twointeg{G}{G}{G}{G};
\\U_x&=&\twointeg{G}{X}{G}{X};
\\j_\mathrm{eh}&=&\twointeg{G}{X}{X}{G}.
\end{eqnarray}
$\alpha$ measures the Cu spin's ability to polarize the ligand, $U_g$ is the Coulomb interaction between electrons in the filled $E_g$ state, $U_x$ is the Coulomb interaction between electron and hole within
one molecule, $j_\mathrm{eh}$ is the electron-hole exchange integral, and $E_{x}$ is energy gap between LUMO and filled $e_g$ state.
From equation~(\ref{eq:excupc}), we can see that the magnitude and sign of $J_1$ depend on the inter-molecule transfer integral $t$. We calculate the matrix element $t$ for polarization transfer by considering the individual hole and electron hoppings among the four states below (Figure~\ref{pic:exciton4}). We find $t=\frac{2t_et_g}{E_x+U_x}$, where $t_{g,e}$ are the single-particle transfer integrals between the filled states and LUMO of different molecules, respectively.

If we consider the contributions from both components of $e_g$ symmetry, the total transfer integral reads
\begin{eqnarray}\label{eq:excitons}
t&=&\frac{t^x+t^y}{u+E_x};
\\t^x&=&t_e^xt_g^x;
\\t^y&=&t_e^yt_g^y;
\\t_e^i&=&\bra{e^{\mathrm{A},\mathrm{LUMO}}_{g,i}}\hat{H}_{core}\ket{e^{\mathrm{B},\mathrm{LUMO}}_{g,i}};
\\t_g^i&=&\bra{e^{\mathrm{A},\mathrm{filled}}_{g,i}}\hat{H}_{core}\ket{e^{\mathrm{B},\mathrm{filled}}_{g,i}},\label{eq:excitons2}
\end{eqnarray}
where $i\in\{x,y\}$, $\hat{H}_{core}$ is the core
Hamiltonian for a Cu(II)Pc dimer. In the present calculations we evaluate $\hat{H}_{core}$ using the Gaussian 98 code \cite{gaussian98}, using the same basis set and exchange-correlation functional described above. The
symbols A,B refer to these two molecules, and $|e^{\mathrm{A,B}}_g\rangle$
refer to the $e_{g}$-symmetry single-molecule ligand states belonging to molecule A
or B.

\begin{figure}[htbp]
\begin{tabular}{c}
\includegraphics[width=7cm,height=5cm]{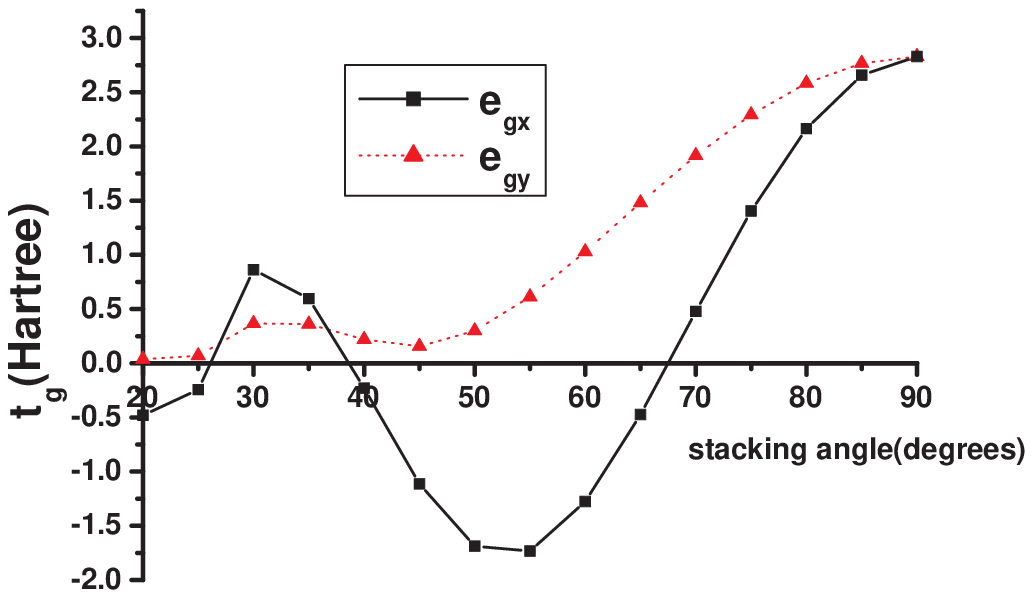}\\
(a)\\
\includegraphics[width=7cm,height=5cm]{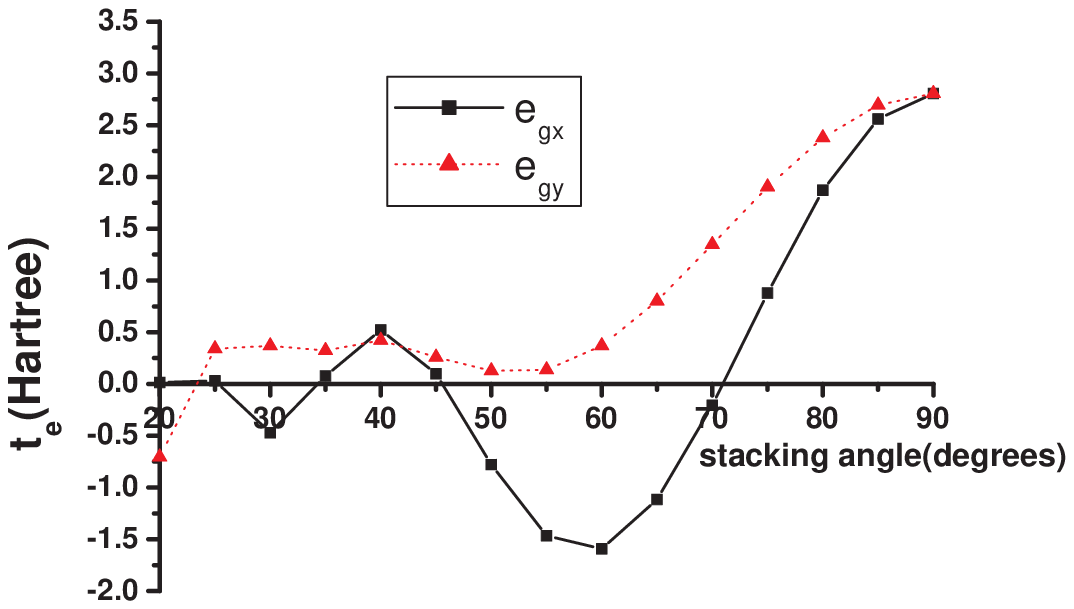}\\
(b)\\
\end{tabular}
\caption{(Color online.) The variation in hopping integrals defined in equations~(\ref{eq:excitons})--(\ref{eq:excitons2}) with stacking angle: (a) $t_g$  for the filled $e_g$ states and (b) $t_e$ for the empty $e_g$ states. In each figure, the solid black curve with square points denotes hopping integrals between $x$-oriented states of the different molecules, and the dashed red curve with triangular points shows the integrals between $y$-oriented states.}\label{pic:tgte}
\end{figure}

We use the single-molecule orbital coefficients of the isolated molecules and the core Hamiltonian for the molecular dimer to calculate the transfer integrals $t_e, t_g$ in the molecular configurations with different stacking angles ($20^\circ-90^\circ$) as shown in Figure~\ref{pic:stacking}. The distance between these two planes is $3.4$ Angstroms \cite{cgb70, hoshino}. In Figure~\ref{pic:tgte}, we show the variation of inter-molecule hopping integrals with stacking angle;  $t_g$ and $t_e$ change both magnitude and sign with stacking angle; this contributes to corresponding changes in the polarization hopping matrix element $t$ and the exchange constant $J_1$.

\begin{figure}[htbp]
\begin{center}
\includegraphics[width=7cm,height=6cm]{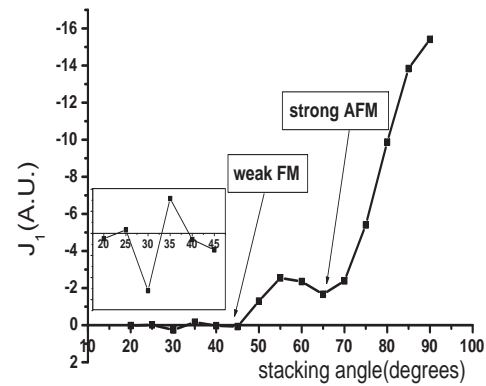}
\caption{Variation of the indirect exchange $J_1$ with the stacking angles shown in Figure~\ref{pic:stacking}; arbitrary units (A.U.) are used.}\label{pic:excofangle}
\end{center}
\end{figure}

In Figure~\ref{pic:excofangle}, we display $t^x+t^y$, which contributes the dependence on stacking angle to $t$ and hence to $J_1$, as a function of stacking angle in the range ($20^{\circ}-90^{\circ}$). When the angle is equal to $45^\circ$, we find weak ferromagnetic (nearly paramagnetic) coupling. When the angle is equal to $65^\circ$, the magnetic interaction is relatively strong anti-ferromagnetic. This calculation qualitatively agrees with the experimental results \cite{sandrine}, though this calculation cannot predict the absolute magnitude of the exchange coupling.

\subsubsection{Mn(II)Pc}
The Mn(II)Pc calculation is more complicated because there are three
unpaired electrons per molecule which occupy $a_{1g}$, and $e_g$
states, so it is necessary to use group theory to simplify the
calculation of the two-electron integrals. By a similar procedure to Cu(II)Pc (the details are shown in  Appendix~\ref{sec:mnpcapp}), we find a weak ferromagnetic interaction when the stacking angle is $45^{\circ}$ but a
relatively strong anti-ferromagnetic interaction for $65^{\circ}$. Unfortunately, even when combined with the super-exchange results for Mn(II)Pc obtained in \S\ref{sec:superexmn} (which always produce anti-ferromagnetic exchange), this result disagrees with the experimental observation of strong ferromagnetic coupling near $\phi=45^\circ$.

\section{Ab initio DFT calculations}\label{sec:dft}
\subsection{Cu(II)Pc}
\begin{figure}[htbp]
\begin{center}
\includegraphics[width=9.5cm,height=6cm]{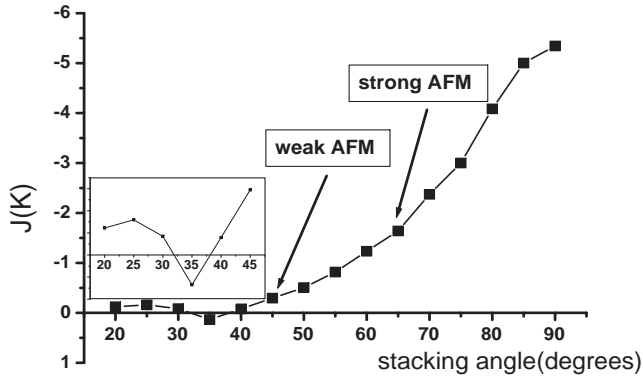}
\caption{The energy difference $J=E_{BS}-E_T$ as a function of stacking angle from $20^\circ$ to $90^\circ$. Notice the qualitative consistency between this figure and Figure~\ref{pic:excofangle}. }\label{pic:dftcomparison}
\end{center}
\end{figure}

We carry out self-consistent calculations of the electronic structure for molecular dimers for the ``+'' structural model at different stacking angles (Figure~\ref{pic:stacking}) by using the Gaussian code with a 6--31G basis set \cite{gaussian98} and the unrestricted B3LYP (UB3LYP) exchange-correlation functional \cite{lyp,b3lyp}. We perform calculations for different stacking angles ranging from $20^\circ$ to $90^\circ$ as shown in Figure~\ref{pic:dftcomparison}; we have tested the convergence of our results with respect to basis set by performing a calculation with a $6\mathrm{-}31\mathrm{+G^{*}}$ basis set (which includes additional polarization functions and diffuse functions) at a single stacking angle ($45^\circ$) and find negligible changes. We compare directly the DFT total energies, and hence calculate the exchange splitting from the difference of the total energies of the broken-symmetry low-spin state and
high-spin state.  For all stacking angles we find it necessary to optimize carefully the occupancy of the Kohn-Sham orbitals in order to
ensure that there is no charge disproportionation between the
molecules; our lowest-energy converged states have Mulliken
charges of approximately +1.00, and nominal spin populations of
$\pm 0.68$, on each Cu atom.  We also need to ensure that the numerical convergence error in the DFT calculations is much smaller than the order of the exchange couplings ($1 \mathrm{K} \sim 10^{-6} \mathrm{Hartree}$); in our calculations, we converge to at least $10^{-9}$ Hartree.  It is encouraging that we find negligible spin contamination in our final Kohn-Sham wave-functions, i.e. the $\langle\hat{S}^2\rangle$ computed for the fictitious non-interacting Kohn-Sham determinants is close to 2.0 for the triplets ($\langle{\hat{S}^2}\rangle=2.0053$) and to 1.0 for
broken-symmetry states ($\langle{\hat{S}^2}\rangle=1.0053$)---note, however, that this is not the same as the expectation value of $\hat{S}^2$ in the true many-body wave function.

\begin{figure}[htbp]
\begin{tabular}{c}
\includegraphics[width=8.51cm,height=7.23cm]{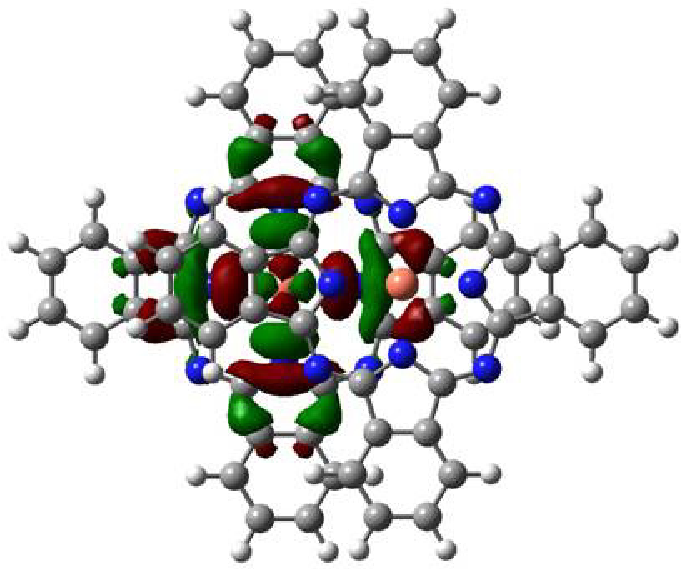}\\
(a)\\
\includegraphics[width=8.11cm,height=6.22cm]{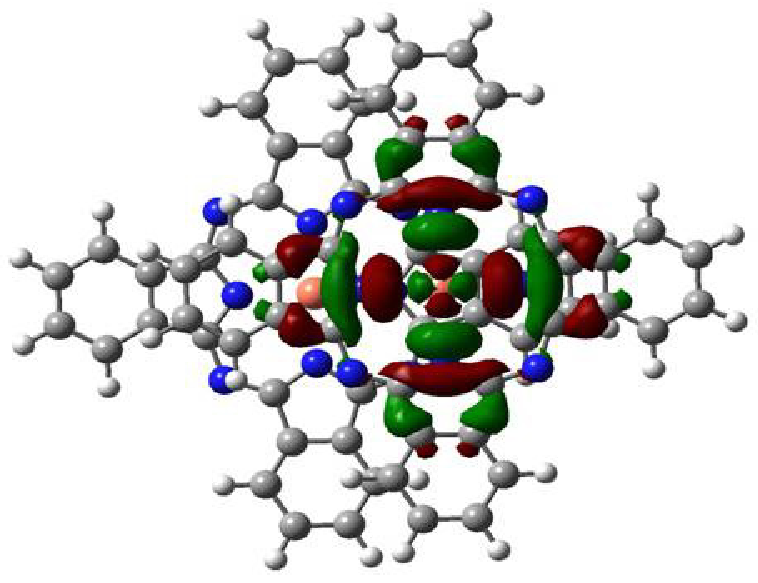}\\
(b)\\
\end{tabular}
\caption{(Color online.) The broken-symmetry orbitals for unpaired electrons in each molecule with spin-up (a) and spin-down (b) from our DFT calculations for Cu(II)Pc dimer (stacking angle $65^\circ$ ). Notice that $b_{1g}$ states are localized on different molecules.}\label{pic:dicupcorbit}
\end{figure}

In the broken-symmetry state, one $b_{1g}$ orbital with spin up is localized on one molecule; the other with spin down on the other molecule as shown in Figure~\ref{pic:dicupcorbit}. Meanwhile, in the triplet state, two $b_{1g}$ orbitals with spin up are localized on both molecules. This is consistent with the DFT calculation of isolated Cu(II)Pc molecule in which localized $b_{1g}$ state carries the unpaired metal electron.

The predicted trend of the exchange
couplings is consistent with perturbation theory calculations shown in Figure~\ref{pic:excofangle}, and in particular shows a strong increase in the coupling as the molecules approach perfect $\pi$ stacking ($\phi=90^\circ$). For the $\beta$ phase, ($\phi=45^\circ$) we have $J=E_{BS}-E_T=-1.1\times 10^{-6}\,\mathrm{Hartree}\approx -0.3 \mathrm{K}$ (See Figure~\ref{pic:dftcomparison}), in agreement with the experimental observation of a nearly paramagnetic state at accessible temperatures, and for $\alpha$-phase ($\phi=65^\circ$) we have $J=E_{BS}-E_T=-5.5\times 10^{-6}\,\mathrm{Hartree}\approx -1.7\,\mathrm{K}$ (see Figure~\ref{pic:dftcomparison}) which gives us a very good agreement with experimental observation $J\sim -1.5\,\mathrm{K}$.

The magnitude of the exchange couplings in Cu(II)Pc is very small: about $10^{-6}$ Hartree, which is right at the edge of the accuracy of the DFT calculation, since there will be errors from the imperfect density functionals and from the finite basis sets as well as the numerical convergence errors discussed above.  However, we can have some confidence in these results for three reasons. First, they agree remarkably well with the magnetization measurements made by the SQUID technique \cite{sandrine}. Second, many of the sources of DFT error could be expected to cancel when we compute the energy difference between systems that are so similar in every respect except their spin orientation.  Third, as discussed above, the results agree with the trends predicted by perturbation theory.

\section{Conclusion and discussion}\label{sec:conclusion}
From perturbative calculations of Cu(II)Pc and Mn(II)Pc we find that
the exchange interaction between two Cu(II)Pc molecules is
dominated by indirect exchange. When the stacking angle is $65^\circ$, the indirect exchange is predicted to be
anti-ferromagnetic, while when the stacking angle is $45^\circ$,
it is very weakly ferromagnetic. Both these
results agree qualitatively with the experimental observations
(see \S\ref{sec:introduction}).

In Mn(II)Pc, by contrast, both super-exchange and indirect exchange contribute. The sign of the
indirect exchange interaction in both cases is dependent on the
sign of inter-molecule electron transfer integrals, and hence varies with stacking angle; however, the
most important terms in the super-exchange are always positive (anti-ferromagnetic).

The main discrepancy with the experiments is in the case of
Mn(II)Pc, where our perturbative calculations do not give the very strong
ferromagnetic interaction which was observed.  This is probably
because the true exchange interaction involves the competition
between super-exchange (always antiferromagnetic) and indirect
exchange (predicted to be once again anti-ferromagnetic at
$65^\circ$, weakly ferromagnetic at $45^\circ$), as well as
possibly other routes.  The different mechanisms involve different
intra-molecular couplings, and so this competition is very
difficult to quantify on the basis of model calculations.

Despite the very different methodology, DFT calculations on Cu(II)Pc produce results that are
remarkably consistent with the perturbation theory. When the angle becomes small, the oscillatory structure of exchange interactions calculated by both perturbation theory and DFT is a signature of the indirect exchange interaction, rather as conventional RKKY oscillations are in a normal metal.
\begin{acknowledgments}
We wish to acknowledge the support of the UK Research Councils
Basic Technology Programme under grant GR/S23506.  We thank Gabriel Aeppli, Sandrine Hertz, Chiranjib Mitra, Marshall Stoneham, Hai Wang, and Dan Wheatley for helpful discussions.
\end{acknowledgments}

\appendix

\section{The indirect exchange for Mn(II)Pc}\label{sec:mnpcapp}
First we need to find the symmetry properties of
the products of pairs of one-electron functions that appear in
equation~\ref{eq:coulombv}.  Here we consider the most complicated case, the product of two
$e_g$ states. Eventually, we will consider the scattering
between filled and empty $e_g$ levels in the molecule, through
interaction with the $e_g$ states of the Mn ion. To do this, we
need the elements of the matrix $X$ such that
\begin{eqnarray}\label{eq:clebschgordan}
X^{-1}[e_{g}\otimes e_{g}]X&=&a_{1g}\oplus b_{1g}\oplus
a_{2g}\oplus b_{2g},
\end{eqnarray}
which are the Clebsch-Gordan coefficients for the product
representation $e_g\otimes e_g$. We can label them as
$X(\alpha,ij)$ where $\alpha$ refers to one of the irreducible
representations appearing on the right of equation
(\ref{eq:clebschgordan}), and $i,j$ label the functions
transforming as $e_g$.  We find
\begin{eqnarray}
X=\bordermatrix{&a_{1g}&a_{2g}&b_{1g}&b_{2g}
             \cr xx&1/\sqrt{2}&0&1/\sqrt{2}&0
             \cr xy&0&1/\sqrt{2}&0&1/\sqrt{2}
             \cr yx&0&-1/\sqrt{2}&0&1/\sqrt{2}
             \cr yy&1/\sqrt{2}&0&-1/\sqrt{2}&0}.
\end{eqnarray}
Because $\frac{1}{\abs{\mathbf{r}-\mathbf{r}^{\prime}}}$ belongs to the identity
representation, we can then rewrite $\hat{v}$ as:
\begin{widetext}
\begin{eqnarray}
\hat{v}&=&\sum_{\sigma,\sigma^{\prime}}\sum_{\alpha}\int{d\mathbf{r}d\mathbf{r}^{\prime}\Psi^{(\alpha)*}\frac{1}
{\abs{\mathbf{r}-\mathbf{r}^{\prime}}}\Psi^{(\alpha)}\sum_{ABDE}X(\alpha,AB)X^{*}(\alpha,DE)}\nonumber
\\&&\times \hat{c}_{A,\sigma}^{\dagger}\hat{c}_{B,\sigma^{\prime}}^{\dagger}\hat{c}_{D,\sigma^{\prime}}\hat{c}_{E,\sigma},
\\&=&\frac{1}{2}\sum_{\alpha}\int{d\mathbf{r}d\mathbf{r}^{\prime}\Psi^{(\alpha)*}\frac{1}
{\abs{\mathbf{r}-\mathbf{r}^{\prime}}}\Psi^{(\alpha)}\sum_{\sigma,\sigma^{\prime}}}\hat{O}^{T}M^{\alpha}\hat{P},
\end{eqnarray}
\end{widetext}

\begin{eqnarray}
\hat{O}=\bordermatrix{&
             \cr&\hat{c}_{x,\sigma}^{\dagger}\hat{c}_{x,\sigma^{\prime}}^{\dagger}
             \cr&\hat{c}_{x,\sigma}^{\dagger}\hat{c}_{y,\sigma^{\prime}}^{\dagger}
             \cr&\hat{c}_{y,\sigma}^{\dagger}\hat{c}_{x,\sigma^{\prime}}^{\dagger}
             \cr&\hat{c}_{y,\sigma}^{\dagger}\hat{c}_{y,\sigma^{\prime}}^{\dagger}}
\end{eqnarray}
\begin{eqnarray}
\hat{P}=\bordermatrix{&
             \cr&\hat{c}_{x,\sigma^{\prime}}\hat{c}_{x,\sigma}
             \cr&\hat{c}_{x,\sigma^{\prime}}\hat{c}_{y,\sigma}
             \cr&\hat{c}_{y,\sigma^{\prime}}\hat{c}_{x,\sigma}
             \cr&\hat{c}_{y,\sigma^{\prime}}\hat{c}_{y,\sigma}}
\end{eqnarray}
\begin{eqnarray}
M^{a_{1g}}&=&\bordermatrix{&
             \cr &1&0&0&1
             \cr &0&0&0&0
             \cr &0&0&0&0
             \cr &1&0&0&1}
\\M^{a_{2g}}&=&\bordermatrix{&
             \cr &0&0&0&0
             \cr &0&1&-1&0
             \cr &0&-1&1&0
             \cr &0&0&0&0}
\end{eqnarray}
\begin{eqnarray}
M^{b_{1g}}&=&\bordermatrix{&
             \cr &1&0&0&-1
             \cr &0&0&0&0
             \cr &0&0&0&0
            \cr &-1&0&0&1}
\\M^{B_{2g}}&=&\bordermatrix{&
             \cr &0&0&0&0
             \cr &0&1&1&0
             \cr &0&1&1&0
             \cr &0&0&0&0}
\end{eqnarray}
\begin{eqnarray}
\Psi^{(A_{1g})}&=&\frac{1}{\sqrt{2}}(\ket{xx}+\ket{yy})
\\\Psi^{(A_{2g})}&=&\frac{1}{\sqrt{2}}(\ket{xy}-\ket{yx})
\\\Psi^{(B_{1g})}&=&\frac{1}{\sqrt{2}}(\ket{xx}-\ket{yy})
\\\Psi^{(B_{2g})}&=&\frac{1}{\sqrt{2}}(\ket{xy}+\ket{yx}).
\end{eqnarray}
We use $X,Y$ to label the ligand $E_{gx}, E_{gy}$ states, and
$x,y$ to label Mn $e_{gx},e_{gy}$ orbitals. Now we introduce
operators which create electron-hole excitations with different
spin symmetries on the Pc:
\begin{eqnarray}
a_{i}^{\dagger(0,0)}&=&a_{i}^{\dagger S}=\frac{1}{\sqrt{2}}(a_{i}^{\dagger\uparrow\downarrow}+a_{i}^{\dagger\downarrow\uparrow});
\\a_{i}^{\dagger(1,0)}&=&a_{i}^{\dagger T}=\frac{1}{\sqrt{2}}(a_{i}^{\dagger\uparrow\downarrow}-a_{i}^{\dagger\downarrow\uparrow});
\\a_{i}^{\dagger\uparrow\downarrow}&=&\hat{c}_{i_{x}\downarrow}^{\dagger}\hat{c}_{i_{g}\downarrow};\qquad
a_{i}^{\dagger\downarrow\uparrow}=\hat{c}_{i_{x}\uparrow}^{\dagger}\hat{c}_{i_{g}\uparrow};
\\a_{i}^{\dagger(1,1)}&=&a_{i}^{\dagger\uparrow\uparrow}=\hat{c}_{i_{x}\uparrow}^{\dagger}\hat{c}_{i_{g}\downarrow};
\\a_{i}^{\dagger(1,-1)}&=&a_{i}^{\dagger\downarrow\downarrow}=\hat{c}_{i_{x}\downarrow}^{\dagger}\hat{c}_{i_{g}\downarrow},
\end{eqnarray}
where $i$ runs over the two orientations of the ligand $e_g$
states ($i=X,\ Y$) and the subscripts $g, x$ of $i$ label the filled ligand $e_g$ states and LUMO ligand $e_g$ states. The following operators characterize the spin degrees of freedom within the subspace where no charge transfer takes place:
\begin{eqnarray}
S_{j}^{z}&=&\frac{1}{2}(n_{j\uparrow}-n_{j\downarrow})
\\S_{j}^{\dagger}&=&\hat{c}_{j\uparrow}^{\dagger}\hat{c}_{j\downarrow}
\\S_{j}^{-}&=&\hat{c}_{j\downarrow}^{\dagger}\hat{c}_{j\uparrow},
\end{eqnarray}
where $j$ runs over all the $e_g$ states of the Mn ion and the
ligand: $j=X_{x},X_{g},Y_{x},Y_{g},x,y$. Using these operators, we
can expand $\hat{v}$ as,
\begin{equation}
\hat{v}=\hat{v}_1+\hat{v}_2+\hat{v}_3+\hbox{spin-independent
terms},
\end{equation}
where
\footnotesize
\begin{eqnarray}
\nonumber\hat{v}_1&=&(a_{X}^{S}+a_{Y}^{S})(2\sqrt{2}(P_3+P_4-P_1/2-P_2/2))
\\\nonumber&&+2\sqrt{2}a_{X}^{T}(P_1S_{x}^{z}+P_2S_{y}^{z})+2\sqrt{2}a_{Y}^{T}(P_2S_{x}^{z}+P_1S_{y}^{z})
\\\nonumber&&+2S_{x}^{\dagger}(P_1a_{X}^{\dagger\downarrow\downarrow}+P_2a_{Y}^{\dagger\downarrow\downarrow})+
2S_{y}^{\dagger}(P_1a_{Y}^{\dagger\downarrow\downarrow}+P_2a_{X}^{\dagger\downarrow\downarrow})
\\&&-2S_{x}^{-}(P_1a_{X}^{\dagger\uparrow\uparrow}+P_2a_{Y}^{\dagger\uparrow\uparrow})
-2S_{y}^{-}(P_1a_{Y}^{\dagger\uparrow\uparrow}+P_2a_{X}^{\dagger\uparrow\uparrow})
\\\nonumber\hat{v}_2&=&2(n_{x\uparrow}[-P_5n_{X_{x}\uparrow}-P_6n_{Y_{x}\uparrow}]
+n_{x\downarrow}[-P_5n_{X_{x}\downarrow}-P_6n_{Y_{x}\downarrow}]
\\\nonumber&&+n_{y\uparrow}[-P_5n_{Y_{x}\uparrow}-P_6n_{X_{x}\uparrow}]
+n_{y\downarrow}[-P_5n_{Y_{x}\downarrow}-P_6n_{X_{x}\downarrow}]
\\\nonumber&&+S_{x}^{\dagger}[-P_5S_{X_{x}}^{-}-P_6S_{Y_{x}}^{-}]
+S_{y}^{\dagger}[-P_5S_{Y_{x}}^{-}-P_6S_{X_{x}}^{-}]
\\&&+S_{x}^{-}[-P_5S_{X_{x}}^{\dagger}-P_6S_{Y_{x}}^{\dagger}]+
S_{y}^{-}[-P_5S_{Y_{x}}^{\dagger}-P_6S_{X_{x}}^{\dagger}])
\\\nonumber\hat{v}_3&=&2(n_{x\uparrow}[-P_5^{\prime}n_{X_{g}\uparrow}-P_6^{\prime}n_{Y_{g}\uparrow}]
+n_{x\downarrow}[-P_5^{\prime}n_{X_{g}\downarrow}-P_6^{\prime}n_{Y_{g}\downarrow}]
\\\nonumber&&+n_{y\uparrow}[-P_5^{\prime}n_{Y_{g}\uparrow}-P_6^{\prime}n_{X_{g}\uparrow}]
+n_{y\downarrow}[-P_5^{\prime}n_{Y_{g}\downarrow}-P_6^{\prime}n_{X_{g}\downarrow}]
\\\nonumber&&+S_{x}^{\dagger}[-P_5^{\prime}S_{X_{g}}^{-}-P_6^{\prime}S_{Y_{g}}^{-}]+
S_{y}^{\dagger}[-P_5^{\prime}S_{Y_{g}}^{-}-P_6^{\prime}S_{X_{g}}^{-}]
\\&&+S_{x}^{-}[-P_5^{\prime}S_{X_{g}}^{\dagger}-P_6^{\prime}S_{Y_{g}}^{\dagger}]+
S_{y}^{-}[-P_5^{\prime}S_{Y_{g}}^{\dagger}-P_6^{\prime}S_{X_{g}}^{\dagger}]),
\end{eqnarray}
\normalsize
and
\begin{eqnarray}
P_1&=&\twointeg{x}{X_x}{X_g}{x}+\twointeg{y}{Y_x}{Y_g}{y}
\\P_2&=&\twointeg{x}{Y_x}{Y_g}{x}+\twointeg{y}{X_x}{X_g}{y}\nonumber
\\P_3&=&\twointeg{X_x}{x}{X_g}{x}+\twointeg{Y_x}{y}{Y_g}{y}\nonumber
\\P_4&=&\twointeg{X_x}{y}{X_g}{y}+\twointeg{Y_x}{x}{Y_g}{x}\nonumber
\\P_5&=&\twointeg{x}{X_x}{X_x}{x}+\twointeg{y}{Y_x}{Y_x}{y}\nonumber
\\P_6&=&\twointeg{x}{Y_x}{Y_x}{x}+\twointeg{y}{X_x}{X_x}{y}\nonumber
\\P_5^{\prime}&=&\twointeg{x}{X_g}{X_g}{x}+\twointeg{y}{Y_g}{Y_g}{y}\nonumber
\\P_6^{\prime}&=&\twointeg{x}{Y_g}{Y_g}{x}+\twointeg{y}{X_g}{X_g}{y}\nonumber.
\end{eqnarray}
Here we can see $\hat{v}_1$ governs the creation of the electron-hole pair, while $\hat{v}_2$ and $\hat{v}_3$ represent the exchange interactions between spins on the Mn ion and on the ligand.
Using this form of $\hat{v}$, we can build the Hamiltonian matrix
for two sets of wave functions: those in which the total
$z$-component of spin on one molecule (Mn plus ligand) is
respectively $+3/2$ and $+1/2$.  We label the individual states as
$\ket{S_{\mathrm{Mn(II)}},S_{\mathrm{Pc}}}$, where the first index is the spin
configuration of Mn ion, and the second is the spin configuration of the ligand
in the $X$ or $Y$ spatial component.
\begin{enumerate}
\item The $+3/2$ states are
$\ket{(3/2),g}$, $\ket{(3/2),(S=0,M=0)}$, $\ket{(3/2),(S=1,M=0)}$,
$\ket{(1/2),(S=1,M=1)}$, and the corresponding matrix is
\begin{eqnarray}
\hat{v}_{\frac{3}{2}}=\bordermatrix{&&&&
             \cr &0&\beta&\frac{3}{2}\alpha&-\sqrt{\frac{3}{2}}\alpha
             \cr &\beta&0&0&\gamma
             \cr &\frac{3}{2}\alpha&0&0&\sqrt{6}\delta
             \cr &-\sqrt{\frac{3}{2}}\alpha&\gamma&\sqrt{6}\delta&0},
\end{eqnarray}where
$\alpha=2\sqrt{2}/3(P_1+P_2),\beta=2\sqrt{2}(P_3+P_4-1/2P_1-1/2P_2),
\gamma=\sqrt{6}/3(P_5^{\prime}+P_6^{\prime}-P_5-P_6),\delta=1/3(-P_5^{\prime}-P_6^{\prime}-P_5-P_6)$.
\item The $+1/2$ states are
$\ket{(1/2),g}$, $\ket{(1/2),(S=0,M=0)}$, $\ket{(1/2),(S=1,M=0)}$,
$\ket{(-1/2),(S=1,M=1)}$, $\ket{(3/2),(S=1,M=-1)}$, and the matrix
of Coulomb interaction $\hat{v}_{\frac{1}{2}}$ is
\begin{eqnarray}
\bordermatrix{&&&&&
             \cr &0&\beta&1/2\alpha&-\sqrt{2}\alpha&\sqrt{3/2}\alpha
             \cr &\beta&0&0&2/\sqrt{3}\gamma&-\gamma
             \cr &1/2\alpha&0&0&2\sqrt{2}\delta&\sqrt{6}\delta
             \cr &-\sqrt{2}\alpha&2/\sqrt{3}\gamma&2\sqrt{2}\delta&0&0
             \cr &\sqrt{3/2}\alpha&-\gamma&\sqrt{6}\delta&0&0}.
\end{eqnarray}
\end{enumerate}

To get the leading terms in the effective
Hamiltonian~\ref{eq:threehalveswiththreehalves}, we need consider
the situation where the total $z$-direction angular momentum on
\textit{both} molecules is
$M_{total}=M_s^{\mathrm{Mn}_A}+M_s^{\mathrm{Mn}_B}+M_s^{\mathrm{Pc}}=2$.
If we restrict ourselves to excitations in which there is only only one electron-hole pair in total
on the two Mn(II)Pc molecules, and suppose it resides in the $X$-symmetry orbitals (the $Y$-symmetry states are completely decoupled in the ``$+$'' model), we are left with a total of $18$ states . These are made up as
follows:

\footnotesize
\begin{eqnarray}
&&(M_s^{\mathrm{Mn}_A}=3/2,M_s^{\mathrm{Mn}_B}=1/2,M_s^{\mathrm{Pc}}=0)\rightarrow
5 \ \mathrm{states}
\\&&(M_s^{\mathrm{Mn}_A}=1/2,M_s^{\mathrm{Mn}_B}=3/2,M_s^{\mathrm{Pc}}=0)\rightarrow
5 \ \mathrm{states}\nonumber
\\&&(M_s^{\mathrm{Mn}_A}=1/2,M_s^{\mathrm{Mn}_B}=1/2,M_s^{\mathrm{Pc}}=1)\rightarrow
2 \ \mathrm{states}\nonumber
\\&&(M_s^{\mathrm{Mn}_A}=3/2,M_s^{\mathrm{Mn}_B}=-1/2,M_s^{\mathrm{Pc}}=0)\rightarrow
2 \ \mathrm{states}\nonumber
\\&&(M_s^{\mathrm{Mn}_A}=-1/2,M_s^{\mathrm{Mn}_B}=3/2,M_s^{\mathrm{Pc}}=0)\rightarrow
2 \ \mathrm{states}\nonumber
\\&&(M_s^{\mathrm{Mn}_A}=3/2,M_s^{\mathrm{Mn}_B}=3/2,M_s^{\mathrm{Pc}}=-1)\rightarrow 2 \ \mathrm{states}\nonumber.
\end{eqnarray}
\normalsize
The perturbation includes both the Coulomb interaction $\hat{v}$
discussed above, and the hopping $t$ which transfers an
electron-hole pair from one Mn(II)Pc to another. We found the
leading term of spin-$\frac{3}{2}$ couplings $J_1$ to be:
\begin{eqnarray}
J_1&=&\frac{3\alpha^2t}{(U_g-U_x-E_x-2j_\mathrm{eh})^2},
\end{eqnarray} where
the definitions of $U_g$, $U_x$, $E_x$, $j_\mathrm{eh}$, $t_e$ and $t_g$ are
the same as those in the Cu(II)Pc calculation.

Considering other situations such as the product of $a_{1g}$ and $e_g$ states gives qualitatively similar results that depend in the same way on the transfer integrals between molecules. Finally, we
include excitations through both components ($X$ and $Y$) of the ligand $e_g$ states, so as in the
Cu(II)Pc calculation we should combine the electron-hole pair transfer integrals to form
$t=t^x+t^y$.

\end{document}